\documentclass{ws-ijmpd}    

\usepackage{epsf}
\usepackage{amssymb}

\begin{document}                                                                                   
\markboth{Adam D. Helfer}{Quantum Nature of Black Holes}  

\catchline{}{}{}{}
  
\title{QUANTUM NATURE OF BLACK HOLES\thanks{This essay received an 
``honorable mention'' in the 
2004 Essay Competition of the Gravity Research
Foundation -- Ed.}} 
\author{ADAM D. HELFER}  
\address{Department of Mathematics\\
University of Missouri\\
Columbia, MO 65211, U.S.A.}

\maketitle

\begin{history}
\received{}
\revised{}
\accepted{}
\end{history}

\begin{abstract}
I reconsider Hawking's analysis of the effects of gravitational collapse on
quantum fields, taking into account interactions between the fields.  The
ultra-high energy vacuum fluctuations, which had been considered to be an
awkward peripheral feature of the analysis, are shown to play a key role.  By
interactions, they can scatter particles to, or create pairs of particle at,
ultra-high energies.  The energies rapidly become so great that quantum gravity
must play a dominant role.  Thus the vicinities of black holes are essentially
quantum-gravitational regimes.
\end{abstract}

\def\scrip{{\cal I}^-}
\def\scrif{{\cal I}^+}

\section{Introduction}

Since black holes are extreme manifestations of general relativity, one
might expect that exotic quantum effects would be amplified in their
vicinities, perhaps providing clues to quantum gravity.  This hope has
not been fulfilled, however, by the commonly accepted analysis of the
effects of gravitational collapse on quantum theory.\cite{1}\cdash\cite{2}  
That analysis
predicts that macroscopic black holes, at least, are essentially
classical objects, subject to only minor quantum corrections:  small
emissions of low-temperature thermal radiation.
This prediction of thermal radiation is very beautiful, and attractive
in its potential to clarify the connection between thermodynamics and
black-hole theory.  

But the analysis leading to this prediction is not without its 
problems.\cite{3} (See Ref.~\refcite{4} for a recent review).
The most striking is
the {\em trans-Planckian problem:  } the model's reliance on
treating vacuum
fluctuations at exponentially increasing energy scales
by conventional physics.  Also the model assumes that interactions
between the quantum fields may be neglected, even (or especially) as
those energies exponentially increase.
While it is
acknowledged that there are as yet no very good responses to these
concerns, I think it is fair to say that the sentiment in the field
has largely been that they are peripheral difficulties which will
somehow be overcome, leaving essentially intact the picture of a
softly thermally radiating black hole.

My aim here is to argue that the correct picture of a black hole is
very different.  The black hole, far from being an essentially classical
object with only minor quantum corrections, has a strong quantum
character which profoundly affects physics in a region of space--time, a region
extending beyond the hole a distance comparable to the size of the hole itself. 
In this region, not merely ``virtual'' vacuum fluctuations, but real
Planck-scale physics must be prevalent.

What I shall show is that taking interactions between quantum fields into
account completely alters the picture drawn by Hawking.  These interactions can
promote vacuum fluctuations
from ``virtual'' to real effects.  We shall find that 
real ultra-energetic effects, such as scattering and pair-production
of particles, take place in the vicinity of the hole.  The energies
are of the same order as those of the vacuum fluctuations.  This means
that they rapidly pass the Planck scale, at which quantum field theory
must break down and quantum gravity must take over.  
The vicinity of a black hole must be a window on quantum gravity.

{\em Conventions and terminology.  } 
We shall use natural units throughout, so $c$,
$G$ and $\hbar$ are unity.  
We shall need to discuss processes with 
energies approaching, but not at, the Planck
energy; we shall refer to these as ultra-energetic.
We shall only consider the case of a spherically symmetric black hole;
its mass will be $M$.

\section{Gravitational collapse and the red-shift}

While the event horizon of a black hole is best known as a ``point of
no return,'' it is a different property which is most important both in
Hawking's analysis and in this paper.  This is that
light rays (and other
fields) passing close to the event horizon are red-shifted, the
shifts increasing exponentially as later and later rays are
considered.

A standard conformal diagram of a spherically symmetric gravitationally
collapsing space--time is shown in Fig.~1.  The collapsing matter is shaded, and
the event horizon is the dashed line.  A distant observer, looking back inwards
towards the hole, would be represented by a point near future null infinity, and
coordinatized by a ``retarded time'' $u$.  Tracing a light ray perceived by this
observer back into space--time, through the origin (where, in the diagram, it
appears to reflect from the left-hand edge) and out towards the distant past, it
arrives at an ``advanced time'' $v=v(u)$.  Since the intervals $du$, $dv$
between crests of successive rays are related by $dv=v'(u)\, du$, the factor
$v'(u)$ is the red-shift suffered by a radial ray passing from the past, through
the collapsing space--time, into the future.

\begin{figure}[th]
\epsfysize=2.5in
\epsfbox{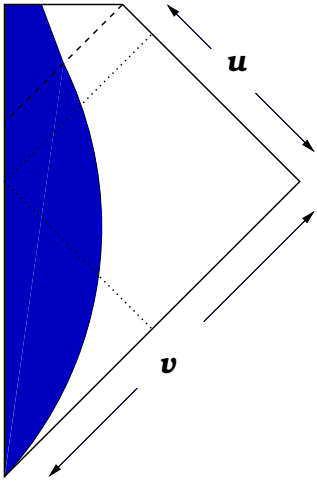}
\caption{A diagram of a black-hole space--time, suppressing angular
variables.  (The left-hand edge is the spatial origin.)
Time increases generally upwards, and lines at $45^\circ$ represent
the paths of radial light rays.  
The scale has been distorted so that the entire
space--time, and some ideal points at infinity, can be represented.  The region
occupied by the collapsing matter is shaded.  The event horizon ${\cal
H}$ is the dashed line; 
the black hole itself is the set of points at and above this.
The dotted line represents a radial light ray beginning at a point on
$\scrip$ (an ideal
set in the past, coordinatized by the advanced time $v$), moving
radially inwards
and passing through the spatial origin (where, in the diagram, it appears to
reflect from the left-hand edge), and escaping to a point
on the ideal set $\scrif$ (coordinatized by $u$).  
}
\end{figure}


It can be shown very generally that the red-shift factor has
asymptotic form
\begin{equation}\label{asympt}
v'(u)\simeq \exp -u/(4M)\quad\hbox{as}\quad u\to +\infty\,
,
\end{equation}
where $M$ is the mass of the hole in geometric units.\cite{5}\cdash\cite{6}
This
exponential decay drives $v'(u)$ to zero very quickly.  For example,
for a solar-mass object, the $e$-folding time is $\simeq 2\times 10^{-5}$ s.  
This exponential asymptotic form plays a key role in Hawking's analysis.

\section{Fluctuations, 
Hawking radiation and the trans-Planckian problem}

Hawking radiation arises because the vacuum fluctuations, which must be present
in the space--time before collapse, are distorted as they propagate through the
collapsing space--time.  In part they are red-shifted by $v'(u)$; but also the
distinction between positive and negative frequencies is distorted.  Since
particle-content is determined by the frequency-decomposition of the operators,
mixing of positive- and negative-frequency terms gives rise to particle
production.  

In the case of the Hawking process, only certain field modes are distorted in
precisely the correct way to produce appreciable particle fluxes.  In fact, a
computation of which particle modes are produced results in the famous
prediction of a thermal spectrum with temperature $T_{\rm H}=1/(8\pi M)$.  

We can give a diagrammatic picture of the Hawking effect, as follows.  While
vacuum fluctuations are not ordinarily drawn as Feynman diagrams (because, in
the absence of a time-dependent external potential, they can be consistently
discarded), they would be represented by closed loops, as in Fig.~2.  In a
gravitationally collapsing space--time, the distortion of the vacuum
fluctuations can be viewed as causing some of the loops to fail to close,
Fig.~3.  Although not expressed in these terms, Hawking's computation is in fact
equivalent to evaluating diagrams like the one in Fig.~3.  While in principle
the virtual portion of the photon line there (that is, everything except the
end-points) could pass anywhere in space--time, the dominant contributions come
from portions like the one shown, where an ultra-high energy vacuum fluctuation
in the past propagates inwards towards the origin, and then outwards, being
distorted into real particles along the way.

\begin{figure}[th]
\epsfxsize=1in
\epsfbox{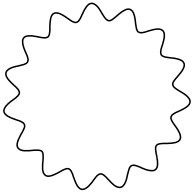}
\caption{A zero-point fluctuation's contribution to the evolution,
represented by a closed virtual photon loop. (Wavy lines represent photons.)
}
\end{figure}

\begin{figure}[th]
\epsfysize=2.5in
\epsfbox{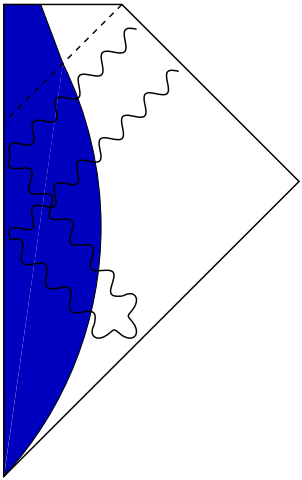}
\caption{One can think of propagation through the time-dependent
gravitationally collapsing space--time as opening some of the 
vacuum fluctuation loops, resulting in real particle production.  
Although, for fixed end-points, one could draw vacuum fluctuation
arcs occupying any portion of space--time, the dominant contributions
to the Hawking process come from ones like that shown here.
Ultra-high frequency fluctuations in the past propagate thought the
collapsing space--time, where they are both red-shifted and distorted
in a way which produces a low flux of real low-energy photons.
}
\end{figure}

Because the outgoing Hawking quanta have characteristic frequencies $\sim \omega
_{\rm H}=1/(8\pi M)$, the vacuum fluctuations which were their precursors must
have had frequencies $\omega _{\rm H}/v'(u)\simeq \omega _{\rm H}\exp +u/(4M)$. 
This is the famous {\em trans-Planckian problem:  } the prediction of mild
thermal radiation requires an appeal to vacuum fluctuations at energy scales
increasing exponentially.  The scales rapidly pass the Planck scale, at which
the model's reliance on ordinary quantum field theory in curved space--time must
break down and quantum gravity must take over.

\section{Effects of interactions}

Hawking's analysis was for free fields.  Let us see what the effects
of including interactions are.
We shall consider for definiteness quantum electrodynamics, but it will be clear
that the basic ideas are more general.

An extraordinary first-order effect occurs when we consider the effect of the
interaction in coupling a charged particle to the vacuum fluctuations, Fig.~4.
In this diagram, the ultra-high energy arc of the vacuum fluctuation interacts
with the charged particle, exchanging energy--momentum.  The remaining portion
of the arc passes through the gravitationally collapsing region, where it
red-shifts (and transmits some of its energy to the hole).  The result is that 
the charged particle has been scattered by an ultra-high energy.  

\begin{figure}[th]
\epsfysize=2.5in
\epsfbox{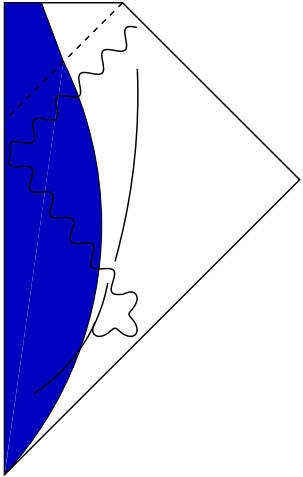}
\caption{A charged particle (solid line)
can interact with an ultra-high frequency 
vacuum fluctuation photon pair (wavy line), 
scattering the charged particle to an ultra-high energy and
producing a Hawking photon.  (This diagram has been drawn to parallel Fig.~3,
and some structure which has conceptual but not literal significance
has been retained.  The bending of the photon line back on itself conveys the
idea of the genesis of an electromagnetic vacuum fluctuation, but is not
literally meaningful.  The crossing of the photon and charged-particle lines is
a consequence of this way of drawing the photon line, and 
likewise not significant.)
}
\end{figure}

In field-theoretic terms, the interaction would be given by $\int A\cdot J\,
d\tau$, where $A$ is the electromagnetic field mode corresponding to the
outgoing Hawking quantum, and $J$ is the current operator.  It is the extension
of $A$ into the past, where it becomes an ultra-high frequency mode mixing both
positive and negative frequencies, which gives rise to the ultra-energetic
scattering.

It is not necessary to have charged particles present initially, however, to
have ultra-energetic effects.  An ultra-energetic vacuum fluctuation can also 
engender pair creation, Fig.~5.

\begin{figure}[th]
\epsfysize=2.5in
\epsfbox{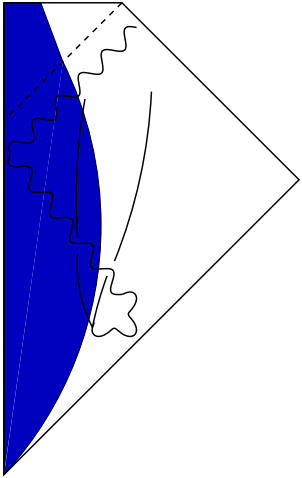}
\caption{An ultra-high energy vacuum electromagnetic fluctuation (wavy
line) can produce an ultra-energetic pair of charged particles
(solid lines).  (Again, the bending-back on itself of the photon line, to
suggest its origin in vacuum fluctuations, has been retained, but this is not of
literal significance.)
}
\end{figure}

(A technical comment.  Feynman diagrams become ambiguous if the definitions of
positive and negative frequencies differ in the in- and out-regions.  
A correct formulation can be inferred by simply considering 
which two-or three-point function is
described by the diagram.)

\section{Conclusion}

We have seen that the presence of interactions will allow ``virtual'' vacuum
fluctuations to produce real physical effects, such as scattering and
pair-creation of particles.  In a gravitationally collapsing space--time, the
result is that ultra-energetic charged particles are produced and are correlated
with the emitted Hawking radiation.  However, the energies increase so rapidly
that, essentially as soon as the hole can be said to have formed, they have
passed the Planck scale.  At that point quantum gravity must enter essentially.

These Planck-scale effects occur, not just in a thin layer around the event
horizon, but in a region of space whose linear size outwards from the black hole
is of the order of the Schwarzschild radius.  This is because, at any point in
space--time, the ultra-energetic effects are essentially determined by the
possibilities of photons being received which have been exponentially
red-shifted by the incipient black hole.  This is essentially determined by the
fraction of the sky occupied by the hole, as seen by an observer at the point in
question, and this fraction will be significant if one is within a few
Schwarzschild radii of the hole.  

We thus predict that the vicinity of a black hole is a region in which
essentially quantum-gravitational, Planck-scale, physics must dominate.


\end{document}